\documentclass[ aps, amsmath,amssymb,reprint,altaffilletter,superscriptaddress,floatfix
]{revtex4-1} 

\usepackage{amsmath} 
\usepackage{amssymb} 
\usepackage{graphicx} 
\usepackage[lofdepth,lotdepth,caption=false]{subfig} 
\usepackage{multirow} 
\usepackage{hyperref} 
\usepackage{placeins} 
\usepackage{color} 
\usepackage{standalone} 

\newcommand{\subfigimg}[3][,]{%
  \setbox1=\hbox{\includegraphics[#1]{#3}} 
  \leavevmode\rlap{\usebox1} 
  \rlap{\hspace*{5pt}\raisebox{\dimexpr\ht1-0\baselineskip}{#2}} 
  \phantom{\usebox1} 
}

\begin{document}
\title{Controlling ferroelectric hysteresis offsets in PbTiO$_{3}$ based superlattices}
\author{Simon Divilov}
\author{Hsiang-Chun Hsing} 
\author{Mohammed Humed Yusuf} 
\author{Anna Gura}
\affiliation{Department of Physics and Astronomy, Stony Brook University, \\Stony Brook, NY 11794-3800 USA}
\author{Joseph A. Garlow}
\author{Myung-Geun Han}
\affiliation{Condensed Matter Physics \& Materials Science, Brookhaven National Laboratory,
Upton, NY, USA 11953}
\author{Massimiliano Stengel}
\affiliation{Institut de Ciencia de Materials de Barcelona (ICMAB-CSIC), Campus UAB, 08193 Bellaterra (Barcelona), Spain}
\affiliation{ICREA-Institucio Catalana de Recerca i Estudis Avançats, 08010 Barcelona, Spain}
\author{John Bonini}
\author{Premala Chandra} 
\author{Karin M. Rabe}
\affiliation{Department of Physics and Astronomy, Rutgers University, NJ USA}
\author{Marivi Fernandez Serra}
 \email{maria.fernandez-serra@stonybrook.edu}
\affiliation{Department of Physics and Astronomy, Stony Brook University, \\Stony Brook, NY 11794-3800 USA}
\affiliation{Institute for Advanced Computational Science, Stony Brook University, \\ Stony Brook, New York 11794-5250, USA}
\author{Matthew Dawber}
 \email{matthew.dawber@stonybrook.edu}
\affiliation{Department of Physics and Astronomy, Stony Brook University, \\Stony Brook, NY 11794-3800 USA}

\begin{abstract}
    Ferroelectric materials are characterized by degenerate ground states with multiple polarization directions. In a ferroelectric capacitor this should manifest as equally favourable up and down polarization states. However, this ideal behavior is rarely observed in ferroelectric thin films and superlattice devices, which generally exhibit a built-in bias which favors one polarization state over the other. Often this polarization asymmetry can be attributed to the electrodes. In this study we examine bias in PbTiO$_3$-based ferroelectric superlattices that is not due to the electrodes, but rather to the nature of the defects that form at the interfaces during growth. Using a combination of experiments and first-principles simulations, we are able to explain the sign of the observed built-in bias and its evolution with composition. Our insights allow us to design devices with zero built-in bias by controlling the composition and periodicity of the superlattices.
\end{abstract}

\maketitle

\section{Introduction}

In an ideal ferroelectric the symmetries of the crystal only allow for the expansion of the free energy in even powers of the polarization, producing two degenerate ground states, one for each polarization direction \cite{Devonshire1954}. 
In the presence of an electric field, the degeneracy breaks due to a linear coupling between the field and polarization, favoring one state over the other depending on the sign of the field, which should produce ferroelectric polarization-field hysteresis loops with two equal but opposite coercive fields.
In theory this should also hold for ferroelectric superlattices that have compositional inversion symmetry \cite{Neaton2003, Johnston2005, Dawber2005}.
However, in practice, electric polarization asymmetry, where one polarization state is preferred over another, or in other words, a built-in bias, is often seen in ferroelectric thin films and superlattices \cite{zhangPRB2014,agarACSNano2015,leeAdvMat2012,zhaoJAP2015,HHwuJAP2013}.
Many contributing factors have been suggested as sources of polarization asymmetry.
Broadly, they can be classified as (i) compositional, eg., asymmetric electrodes, polarization-strain gradient
coupling \cite{zhangPRB2014,KarthikPRB2013,annurev-matsci-071312-121634,10.1142/9789814719322_0002}, chemical and/or compositional
gradients \cite{agarACSNano2015}, and asymmetric interfaces within the superlattice \cite{CalloriPRL2012,PhysRevB.80.224110,PhysRevLett.101.087601}, or (ii) related to defects such as Pb vacancies \cite{gaoNatcomm2011}, oxygen vacancies \cite{gaoNatcomm2011}, Pb-O vacancy complexes \cite{Scott1991}, or other defect dipoles \cite{HHwuJAP2013}.
It is typically difficult to decisively identify the source
of polarization asymmetry in any given material system, and concrete solutions to control and tune this built-in bias are lacking.
In artificially layered superlattices variables such as the chemical composition, the number of interfaces, and even the inversion symmetry at these interfaces \cite{Callori2012} can be precisely controlled, providing an  excellent opportunity to systematically study the built-in bias.

\begin{figure}
\includegraphics[width=8.5cm]{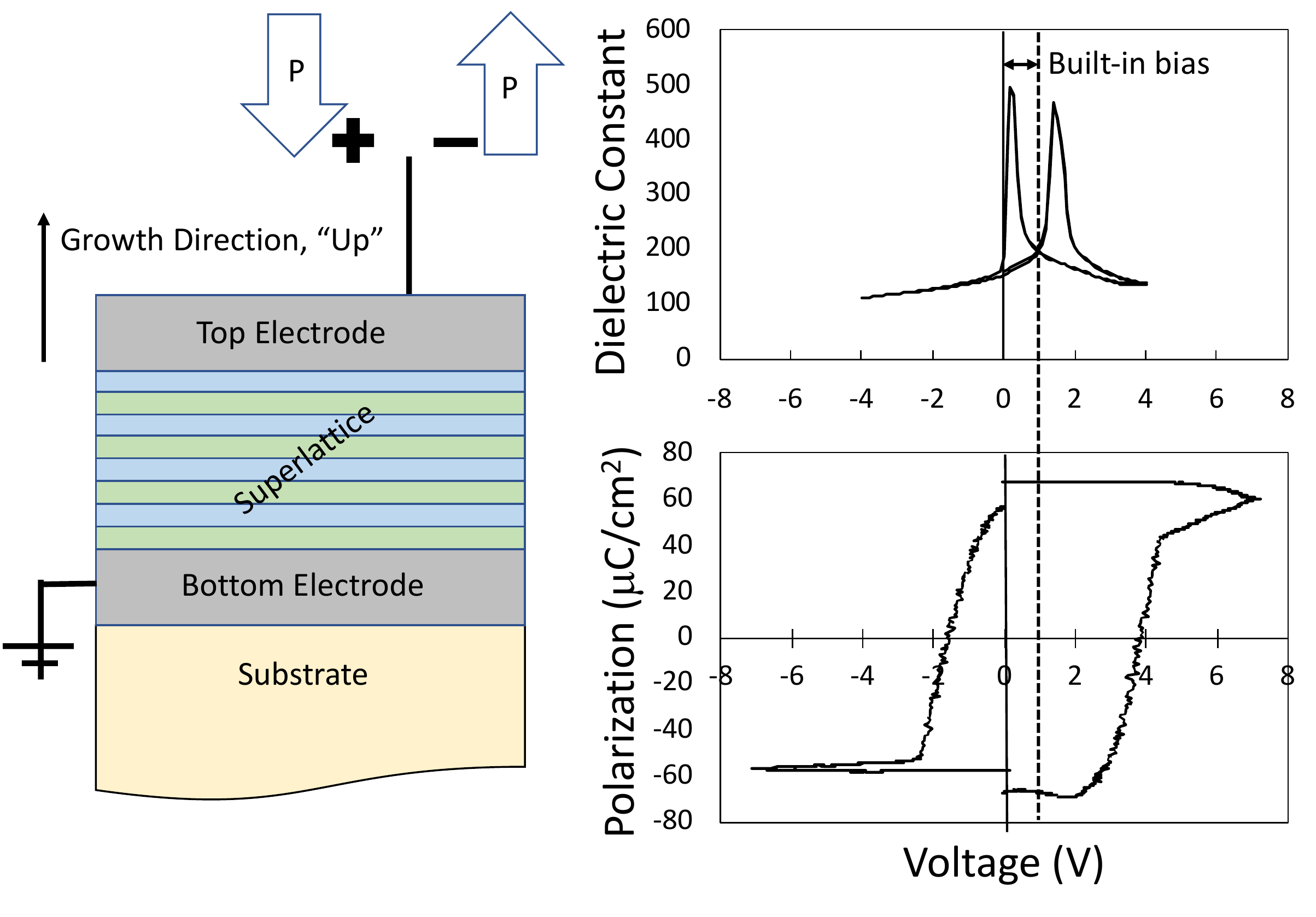}
\caption{Definition of up/down direction in the superlattice samples. Up is considered to be away from the substrate and corresponds to growth direction. Electrical voltage is applied to the top electrode with respect to the grounded bottom electrode. A positive voltage favors down polarization, negative voltage prefers up. Definition of built-in bias for this work as the offset of the Capacitance Voltage loop from zero volts. Experimental data measured in a PTO/STO superlattice.
}
\label{fig:definitions}
\end{figure}

In this paper, we study in detail the origins of this bias in PbTiO\textsubscript{3} based superlattice structures. 
In particular we focus on PbTiO\textsubscript{3}/SrTiO\textsubscript{3} (PTO/STO) and PbTiO\textsubscript{3}/SrRuO\textsubscript{3} (PTO/SRO)
systems.
Our experiments show that the
built-in bias in PTO/STO systems is systematically different from that in PTO/SRO superlattices. 
In order to understand the differences, we gain insight into the physical mechanisms underlying this phenomenon by performing a detailed first-principles study on the role of dipolar defects in these two systems. 
We find that the electrostatic interactions induced by Pb-O divacancies in these superlattice systems
can explain the experimental observations.
To  demonstrate the value of our findings, as well as to confirm the theoretical predictions, we design a system where we take advantage of the presence of Pb-O divacancies to engineer a superlattice with no built-in bias.

\section{Experimental Results and discussion}
\subsection{Composition dependence of built in bias}

Epitaxial growth of n\textsubscript{1}PbTiO\textsubscript{3} (u.c.)/n\textsubscript{2}SrTiO\textsubscript{3}
(u.c.) and n\textsubscript{1}PbTiO\textsubscript{3} (u.c.) /n\textsubscript{3}SrRuO\textsubscript{3}
(u.c.) superlattices was achieved using an off-axis RF magnetron sputtering
deposition system on (001) TiO\textsubscript{2} terminated SrTiO\textsubscript{3}
substrates. 
Here we use the commonly used notation where  n\textsubscript{1} /n\textsubscript{2} refers to the thickness of the layers within the superlattice bilayer repeast in unit cells. The total superlattice film thickness for all samples in this paper is $\approx$100nm. 
A key parameter useful in describing the evolution of properties in these superlattice systems is the PTO volume fraction $(\frac{n_{1}}{n_{1}+n_{2}})$.

Experimentally, electrical properties of ferroelectrics are measured using a variety of techniques and built-in bias can be defined in various ways. 
A key measurement is the polarization-voltage hysteresis loop. This can be done through a continuous sweep of the voltage from zero, to a maximum field in one direction, then to the same field in the opposite direction, before bringing the field back to zero. During this sweep the current flowing can be integrated to give the switched polarization. This assumes that the current is dominated by switching current, whereas in practice there can be substantial dielectric charging and leakage currents. These additional components can be removed using an PUND pulse switching approach, leaving pure switching polarization. The loop shown in Figure. \ref{fig:definitions}  is a result obtained using the PUND approach.
Another important measurement is the capacitance-voltage measurement. Here capacitance is measured using a small AC field superimposed on a larger DC voltage, which is swept around in much the same way as in the polarization-voltage hysteresis measurement, though usually significantly slower.
Empirically, we have found that the center point of the polarization-voltage hysteresis loop and the crossing point in capacitance-voltage loops are closely aligned and represent a useful measurement of the built-in-bias of a sample.
However, in general it is easier to clearly identify the latter. 
Hence experimental values of the built-in bias reported in this paper are obtained from 
capacitance-voltage loop data, as shown in Fig \ref{fig:definitions}.

\begin{figure*}
    \hspace*{-25.0pt}%
    \subfloat{
        \subfigimg[width=0.33\linewidth]{\textbf{a}}{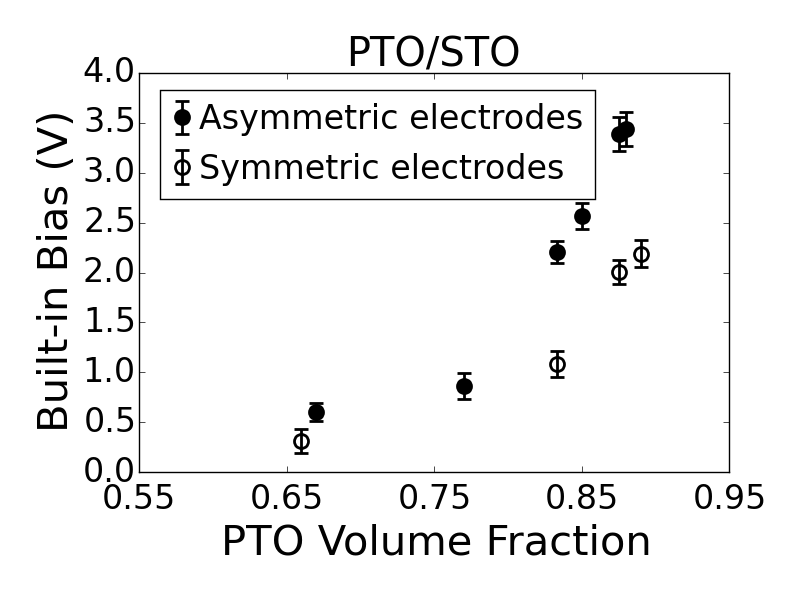}%
        }
        \hspace*{-20.0pt}%
    \subfloat{
        \subfigimg[width=0.33\linewidth]{\textbf{b}}{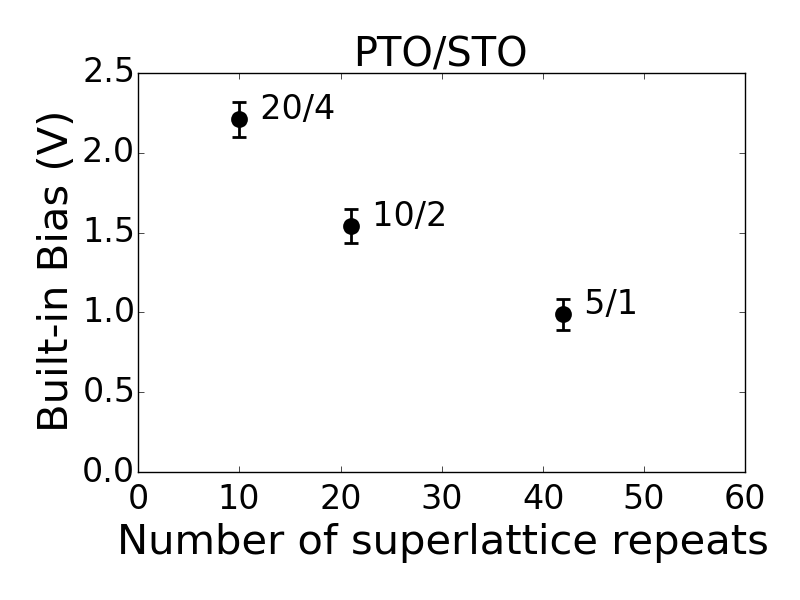}%
        }
        \hspace*{-20.0pt}%
    \subfloat{
        \subfigimg[width=0.33\linewidth]{\textbf{c}}{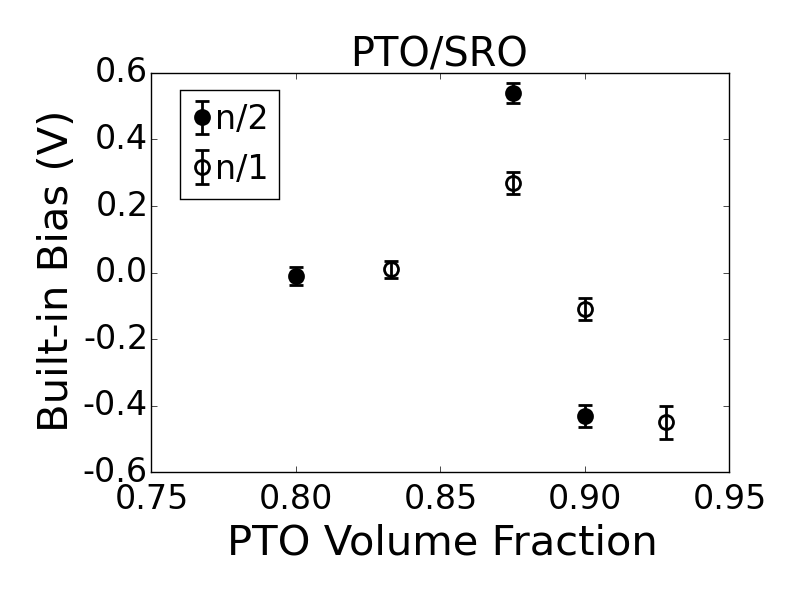}%
        } 
        \hspace*{-20.0pt}%
    \caption{(a) The scaling of the built-in bias in the PTO/STO superlattice system as a function of the PTO Vvolume fraction. (b) Bias dependence on the number of superlattice repeats at a fixed volume fraction in PTO/STO superlattices, and (c) bias as function of volume fraction in PTO/SRO superlattices}
    \label{datasummary}
\end{figure*}

The first trend that we identify experimentally, shown in Fig. \ref{datasummary} (b), is  that the built-in bias rapidly increases as a function of PTO
volume fraction both for samples with asymmetric electodes (SRO-PTO/STO-Pd) as
well as symmetric electrodes (SRO-PTO/STO-SRO).
To investigate the role that interfaces play in the built-in bias of our
superlattice thin films, we prepared a series of superlattice
samples with the same PbTiO\textsubscript{3} volume fraction (0.83)  but
different bilayer thickness, and, as a result, a different number of interfaces.
To minimize bias due to asymmetric top and bottom electrodes in this series, 30nm top and bottom SrRuO\textsubscript{3} layers were deposited in situ followed by photolithography and Ar gas dry etch to pattern the top SrRuO\textsubscript{3}.

The samples prepared were a 20/4 PTO/STO, a 10/2 PTO/STO and a 5/1
PTO/STO composed of 10, 21 and 42 bilayer repeats respectively.
Naively, one can imagine that if the interface is the main source of asymmetry in
the superlattice, the more interfaces that are present the larger the built-in bias should be.
However, as shown in Fig. \ref{datasummary} (b) we found that the built-in bias actually decreased with the increase of interfaces in the superlattice.
As a contrast to PTO/STO superlattices we grew a series of  PTO/SRO superlattices,
that have an additional compositional asymmetry at the interface\cite{CalloriPRL2012}.
This asymmetry occurs because PTO/SRO superlattices have variation in both the A and B site cations, unlike PTO/STO superlattices where variation is confined to the A site, which does not break compositional inversion symmetry.
This compositional breaking of inversion symmetry manifests itself as a built-in bias that occurs even if the samples are free of defects and have symmetric top and bottom electrodes.
However, experimentally we find that the bias offset in these PTO/SRO
samples is actually  significantly smaller than in the PTO/STO system. Perhaps the most interesting feature of the built-in bias in PTO/SRO is that we observe a transition from positive to negative bias in samples with PTO volume fractions of 0.9 and above, in stark contrast to the PTO/STO samples where a positive built-in bias of several volts is observed in samples with these PTO volume fractions.
Our previous work \cite{Callori2012} showed that that PTO/SRO superlattices
with lower PTO volume fractions will display more asymmetric
ferroelectric properties, ie. larger built-in biases, and that the effects of this asymmetry should decrease with increasing PTO volume fractions.
We might then expect that PTO/SRO samples with high PTO volume fraction will display characteristics similar to those observed in PTO/STO superlattices, so the observation of an opposite bias sign in these samples is somewhat surprising. To understand this result we need to look deeply at the origin of bias in these systems.


\subsection{Oxygen vacancies as a source of built-in-bias}

The most common defects in these materials are oxygen vacancies, and hence it is natural to explore whether they can explain the observed experimental behavior.

In order to explore the potential impact of oxygen vacancies on the built-in
bias we performed two sets of annealing experiments.
The first experiment involved annealing a 10/2 PTO/STO superlattice at a relatively low temperature of
400\textsuperscript{o}C.
The annealing time and annealing environment was modified between experiments.
X-ray diffraction $2\theta-\theta$ measurements (XRD) performed before and after annealing (shown in Supplemental Fig S1(a)) show that superlattice
peaks associated with periodic layering remain sharp with high intensity after annealing,  indicating that
no significant structural change occurs at 400\textsuperscript{o}C.
When annealed at this temperature in vacuum, the bias did not change.
However, after annealing in air for 2 hours, the bias dropped from around 1.02 V to 0.83V
and remained at that level after continued annealing in air up to 12hrs. 
This observation is most likely due to a decreased oxygen vacancy concentration in the sample, as one would expect after annealing in air.

The second annealing experiment we performed measured the bias after annealing in vacuum at higher
temperature. 
The experiment was done on a 5/1 PTO/STO superlattice, which was selected due to the 1 unit cell STO layer which we anticipated would be most damaged by the high temperature anneal. 
This time, after annealing in vacuum at 600$^{\circ}$ C, in $2\theta-\theta$ measurements (XRD) performed before and after annealing (shown in Supplemental Fig S1(b)) the superlattice peaks vanished.   
This indicates that the layered structure in the superlattice 
underwent significant disruption, in which the high temperature treatment allows inter-diffusion of Pb and Sr into a more favorable disordered
configuration\cite{Cooper07}, akin to a solid solution, rather than the ordered layered structure that was imposed by the growth process. 
However, even with this degradation of the superlattice structure, the
polarization, as well as the built-in bias
do not change.
No oxygen was introduced during the annealing
treatment in vacuum, and hence the total number of vacancies in the sample remains
the same. 
Unlike the Pb and Sr atoms at the superlattice interfaces, the vacancies do not appear to redistribute under temperature.

\begin{figure}
\begin{centering}
\includegraphics[width=8.5cm]{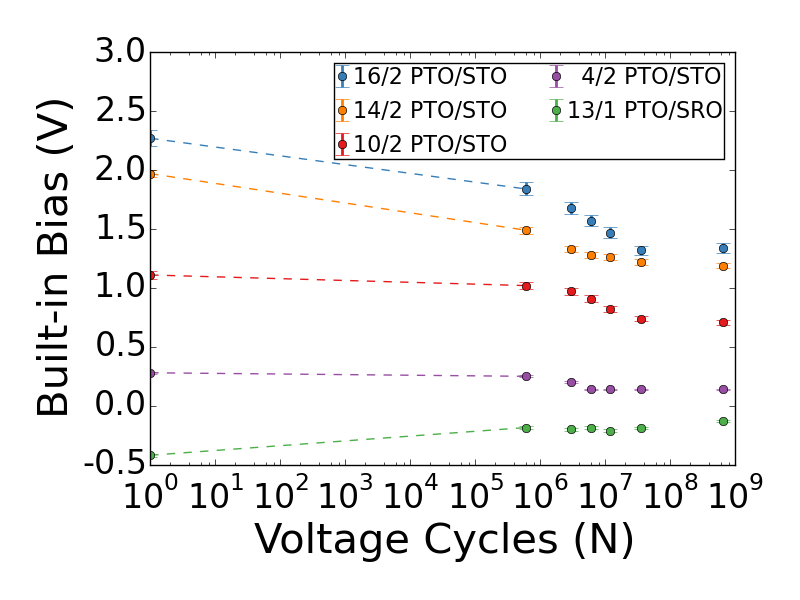}
\par\end{centering}
\caption[Cycling test to explore the effect of oxygen vacancies.]{ The built-in bias as a function of the number of voltage cycles for several
superlattice samples. The bias offset decreases as the number of voltage
cycles increases}
\label{annealfatigue}
\end{figure}

On the other hand, oxygen vacancies can be redistributed through repetitive electrical
cycling\cite{scott2000oxygen,dawber2000model}.
We drove such a redistribution  by applying a 10kHz 4V AC  signal (larger than the coercive field) to 
our samples. 
After the electrical cycling we imposed a
10 minute delay time for the sample to settle before performing
a capacitance butterfly loop measurement to obtain the built-in bias.
Results are presented in Fig.~\ref{annealfatigue}, the magnitude of the built-in bias decreases with the number of voltage
cycles for both the PTO/STO and PTO/SRO samples.
These experiments suggest that defects related to oxygen vacancies 
are clearly implicated in the built-in bias.
In the following sections we explore mechanisms that can explicitly couple
the presence of oxygen vacancies to the built-in-bias.

\subsection{PbO divacancies in superlattices}

The study of oxygen vacancies in oxide perovskites continues to be an active
research field \cite{PhysRevResearch.2.023313}.
Oxygen vacancies in their simplest form can be considered as charged point defects.
However, a uniform distribution of point charges does not produce an electric field and point defects will only lead to built-in-bias when there is a non-uniform distribution, e.g. a local gradient, of them.
Charged point defects can encourage the formation of tail-to-tail or head-to-head domains\cite{Park1998,PhysRevB.94.100101}, but these also do not create a built-in bias
due to their intrinsic symmetry.
On the other hand ordered dipolar defects naturally lead to one polarization state being preferred over the other which would lead to a built-in bias \cite{Arlt1988, Scott1991}.
In PTO, both Pb and O vacancies have been shown to be stable point defects \cite{Zhang2006} and can readily form in the growth of PTO thin films. 
Taking all these into consideration, in this work we choose to model nearest neighbor Pb-O divacancies as the most probable sources of bias.
This choice also reflects that they are found to be lower in energy than next-nearest neighbor divacancies \cite{Pykk2000,Cockayne2004}.
As we will see, the excellent correlation between computational
and experimental results indicates that our choice is well conceived.
The PbO divacancies we consider in the thin film and superlattice systems are formed by removing a Pb atom from a PbO plane and the nearest O atom from the adjacent TiO$_2$ plane.
The dipole moment vector of this divacancy along the [101] direction points to the oxygen vacancy site.
Its out of plane projection results in the up(down) dipole responsible for the observed bias,
as shown in Fig~\ref{fig:stosro_vacloc}.

A priori, it might be considered that PbO divacancies are equally likely to form as either up or down defects, as long as there is no external driving force to break the up/down symmetry.
Under these circumstances, PbO up/down divacancies would form with equal probability, resulting in no net effect on the observed bias.
\begin{figure}
    \centering
    \includegraphics[width=8.5cm]{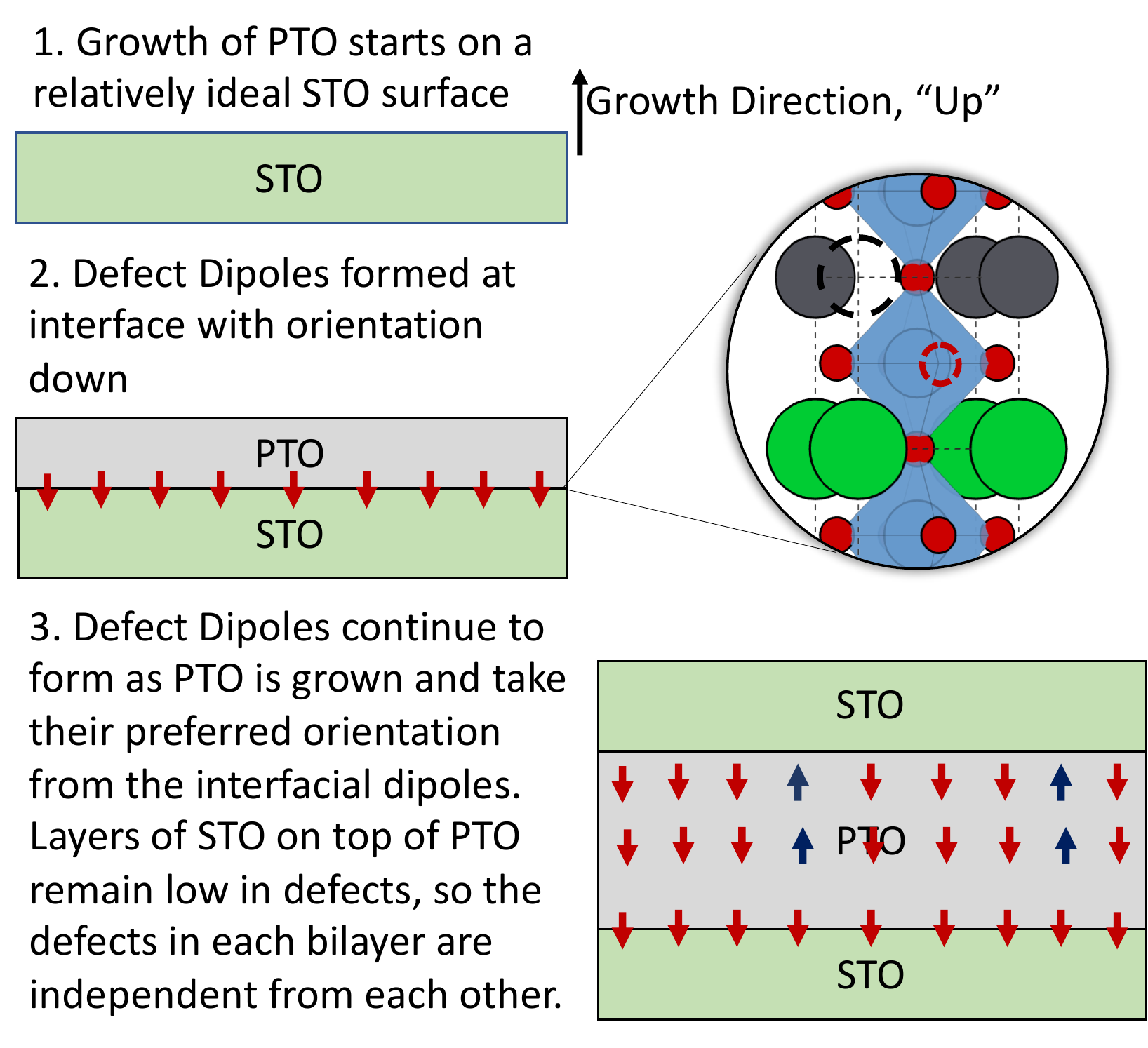}
    \caption{Process through which a preferred orientation of the defect dipoles is developed in the growth of PTO/STO superlattices.}
    \label{fig:PTOSTOmeach}
\end{figure}
However, in our experiments there is an naturally preferred asymmetry introduced by the layer-by-layer growth process. 
This is shown in Fig. \ref{fig:PTOSTOmeach}.
The superlattices are always grown on top of a STO substrate.
Because of this, in PTO/STO superlattices, the first grown interface is STO(bottom)/PTO(top).
Bulk PTO has a larger lattice constant than bulk STO.
Hence the in-plane strain in PTO on STO is compressive at standard deposition temperatures where PTO is cubic\cite{liu2020role,doi:10.1063/1.4939790}.
%
%
The compressive strain on PTO can favor vacancy formation to enable release of the lattice mismatch-induced stress. 
This is also supported by the higher volatility of Pb, which favors the
subsequent formation of O vacancies to ensure charge neutrality.
%
%
This implies that there will be a preferred interface where the divacancies form with a fixed direction of the dipole moment.
In particular, Pb vacancies will be formed at the first grown PbO layer, with O vacancies formed
at the TiO$_2$ layer underneath, with the dipole moment pointing down towards the substrate.
As divacancies continue to form within the bulk of growing PTO layers they will tend to align themselves with the existing divacancies that formed at the interface layer and the collective alignment of the divacancy dipole moments would lead to the experimentally observed bias, hence explaining the observation that the built-in bias increases with increasing PTO volume fraction.
While it is not possible to directly capture this growth induced asymmetry in our simulations, we postulate that it is this phenomenon which allows the divacancies to preferably form in one direction, becoming a significant source of the bias in our samples. 
In the following sections we present the modelling of these divacancies, with results that qualitatively and quantitatively support the experimental observations.
\section{Modeling and Discussion}
\subsection{Methods}

Our simulations of Pb-O divacancies involve density functional theory (DFT) within the local density approximation (LDA) using both numerical atomic orbitals, as implemented in \textsc{siesta} \cite{Soler2002},  and augmented plane-waves as implemented in \textsc{lautrec}.
For \textsc{siesta}, we used norm-conserving pseudopotentials generated using the Trouiller-Martins\cite{Troullier1991} scheme.
The exchange-correlation functional used was PZ\cite{Perdew1981}.
The electrons treated explicitly with a single-$\zeta$ were 5d$^{10}$ (Pb), 4s$^{2}$4p$^{6}$ (Sr, Ru) and 3s$^{2}$3p$^{6}$ (Ti).
The electrons treated with a double-$\zeta$ were 6p$^{2}$ (Pb), 5s$^{2}$ (Sr), 4s$^{2}$3d$^{2}$ (Ti), 5s$^{1}$4d$^{7}$ (Ru) and 2s$^{2}$2p$^{4}$ (O).
The Brillouin zone was sampled using a $6\times6\times1$ Monkhorst-Pack \cite{Monkhorst1976} mesh with a plane-wave- equivalent cutoff energy of 400 Ry.
%
%
The force and pressure tolerances were 0.04 eV/\AA{} and 0.0006 eV/\AA$^{3}$, respectively.
For \textsc{lautrec}, we used the projector augmented wave (PAW) scheme \cite{Blchl1994}.
The exchange-correlation functional used was PW92\cite{Perdew1992} which is equivalent to PZ because they are fit to the same Ceperley-Alder data.
The valence electrons explicitly treated for Pb, Sr, Ti, Ru and O were 6s$^{2}$5d$^{10}$6p$^{2}$, 4s$^{2}$4p$^{6}$5s$^{2}$, 3s$^{2}$3p$^{6}$4s$^{2}$3d$^{2}$, 4s$^{2}$4p$^{6}$5s$^{1}$4d$^{7}$ and 2s$^{2}$2p$^{4}$, respectively.
The Brillouin zone was sampled using a $2\times2\times1$ Monkhorst-Pack mesh with a plane wave cutoff energy of 40 Ry.
%

\subsection{PTO/STO superlattices}

\begin{figure}
\includegraphics[width=8.5cm]{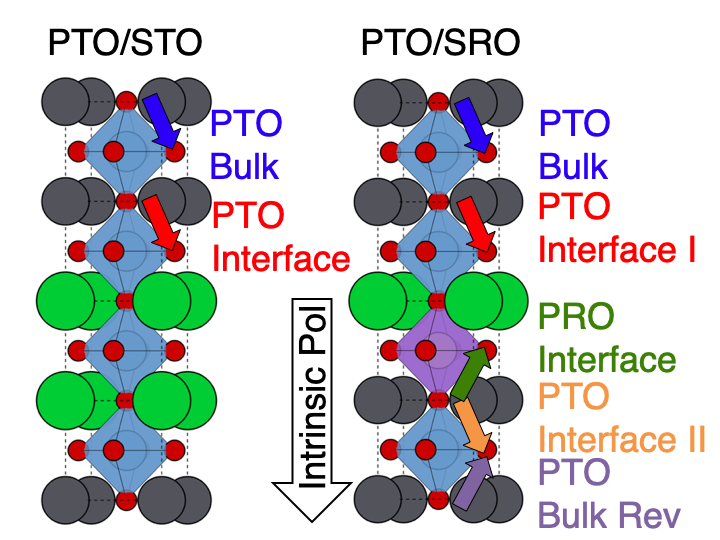}
\caption{Location of the distinct types of Pb-O divacancies and their dipole moments near the interface in the PTO/STO and PTO/SRO superlattice. Grey and red spheres represent Pb and O ions, respectively. Blue and purple octahedra are centered on Ti and Ru ions, respectively.}
\label{fig:stosro_vacloc}
\end{figure}

\begin{figure*}
    \hspace*{-25.0pt}%
    \subfloat{
        \subfigimg[width=0.33\linewidth]{\textbf{a}}{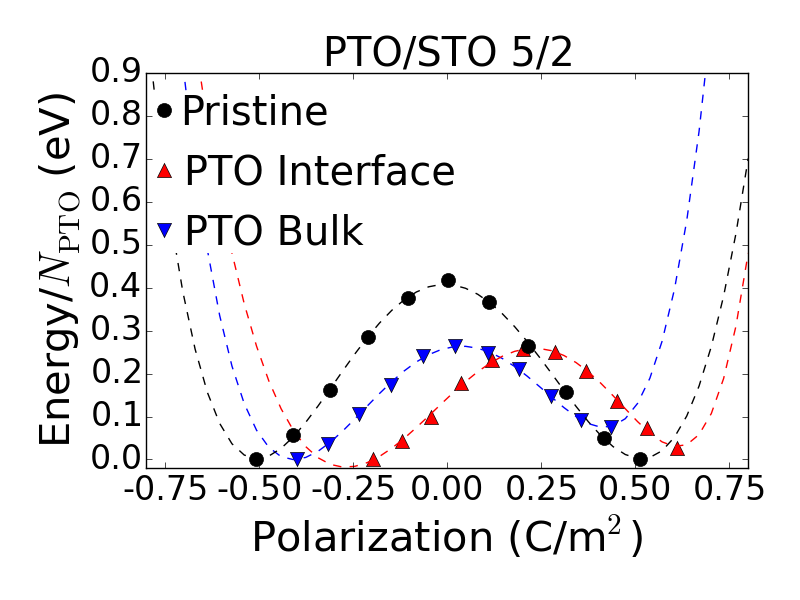}%
        \label{fig:pto5sto2_dw}
        }
        \hspace*{-20.0pt}%
    \subfloat{
        \subfigimg[width=0.33\linewidth]{\textbf{b}}{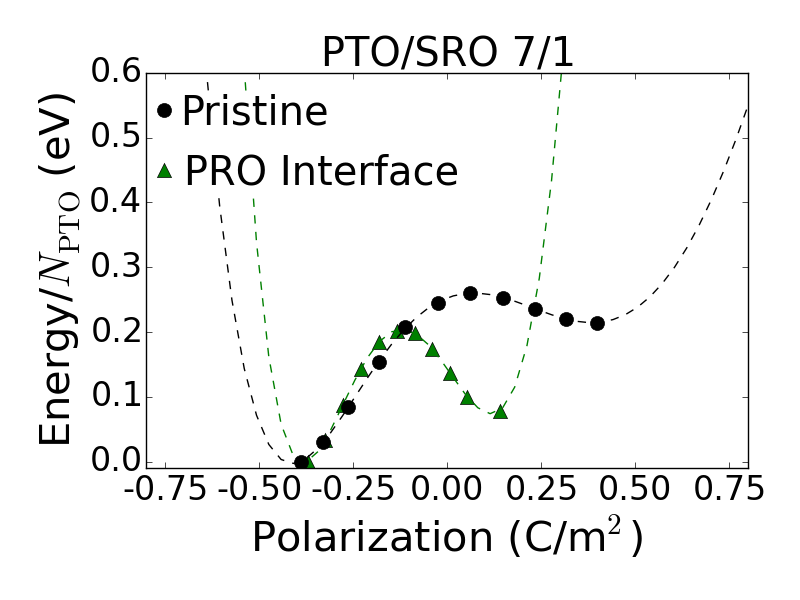}%
        \label{fig:pto7sro1_dw}
        }
        \hspace*{-20.0pt}%
    \subfloat{
        \subfigimg[width=0.33\linewidth]{\textbf{c}}{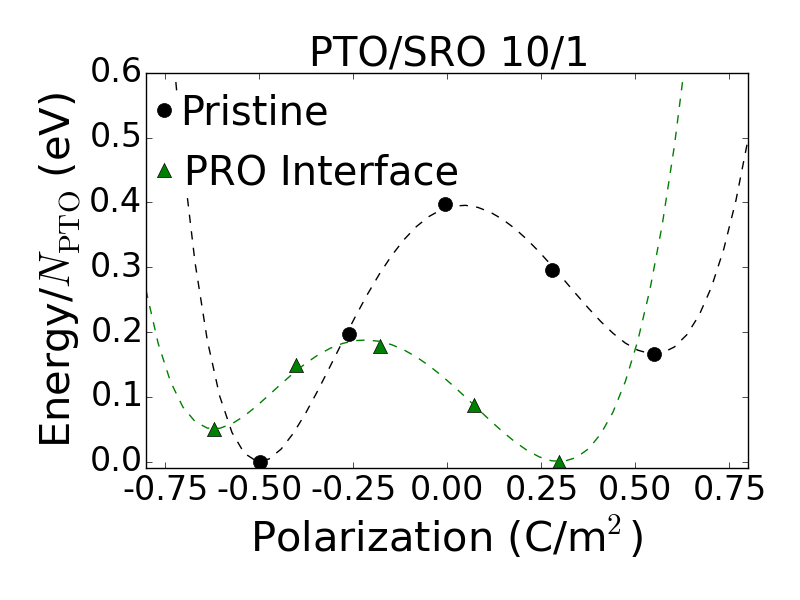}%
        \label{fig:pto10sro1_dw}
        } 
        \hspace*{-20.0pt}%
    \caption{Double well potentials of (a) PTO/STO 5/2, (b) PTO/SRO 7/1 and (c) PTO/SRO 10/1 superlattices. The intermediate points between the two local minima are calculated by interpolating the atomic positions. The plots have all been shifted to have the same zero of energy, fixed to the ground state of the system, for ease of comparison.
    $N$\textsubscript{PTO} is the number of PTO layers in the supercell.
    The dashed lines are best fits to a sixth order polynomial and each curve is plotted relative to its lowest energy state.}
    \label{fig:double_wells}
\end{figure*}

Two distinct types of Pb-O divacancy sites exist in PTO/STO superlattices, as shown in the left schematic of Fig. \ref{fig:stosro_vacloc}. 
We first investigated the energy landscape of a single PbO divacancy in a (PTO)\textsubscript{5}/(STO)\textsubscript{2} superlattice with an in-plane $2\sqrt{2}\times2\sqrt{2}$ supercell.
In particular, we compute the stability of interfacial and bulk divacancies as a function of
the polarization state of PTO. 
To simulate growth on a STO substrate, the in-plane lattice constant was constrained to the STO equilibrium lattice constant, determined from first principles, to be 3.87 \AA.

As previously stated, the divacancies we consider are formed by removing a Pb atom in a PbO plane and an O atom in the adjacent TiO\textsubscript{2} plane. 
Divacancies in bulk PTO are positioned in the center of the PTO layer, as far away as possible from the interface.
The dipole moment of the divacancy has a non-negligible in-plane component.
However, the experimentally applied electric field on the system is only in the [001] direction and therefore the in-plane component does not contribute to the coupling of the out-of-plane polarization with the electric field.
The details of calculating the out-of-plane polarization can be found in Supplementary Section SII.\cite{PhysRevB.73.075121,PhysRevB.75.205121}
We initialize our superlattice in the $P4mm$ space group, which naturally gives rise to polar distortions along the [001] direction.
This is achieved through distorting the TiO\textsubscript{2} planes by displacing the Ti atom above (below) and the O atoms below (above) the planar center, which corresponds to the system being polarized up (down).
 Relaxation of the system, in general, leads to two metastable polarization states of opposite sign.
 
For both bulk and interfacial divacancies, the most stable of these two states in each case aligns the bulk polarization with the dipole moment of the divacancy, as shown in Fig. \ref{fig:pto5sto2_dw}.
In addition, we found that the most stable location of the divacancy is in the interface between the PTO and STO, with an energy 0.04 ev lower than that of the bulk divacancy.
%
%
These results are largely what would be expected from electrostatic considerations.
It is interesting to notice that for interfacial defects, the double well potential is shifted horizontally by a significant amount.
However, it can be shown that this horizontal polarization shift has no effect in the experimental measurement of hysteresis (see Supplemental Fig. S2), nor causes any bias.
This is because it is only the difference between the two polarization states that experimentally is measured through the integrated current as the system's polarization is switched.
On the other hand, as shown in our theoretical modeling of hysteresis loops in the Supplemental Fig. S2, the difference in energy between the two minima of the double well (vertical asymmetry) is proportional
to the bias.

Experimental results shown in Fig~\ref{datasummary}(b), indicate that the
bias does not increase with increasing number of interfaces. 
On the contrary, for the superlattices considered,
the bias decreases roughly linearly, by more than a factor of two, as the number of interfaces increases. 
Our results support and explain this observation, because indeed the bulk
divacancies have a larger vertical asymmetry than the interface divacancies, as seen in Fig.~\ref{fig:pto5sto2_dw}.

Putting together the theoretical results and the symmetry breaking imposed by the growth process, 
we can conclude that the role of the interfaces is to break the symmetry within the bulk by orienting the bulk defect dipoles parallel to the interface defect dipoles.
Effectively, the interfaces polarize the bulk.
For a constant PTO/STO volume fraction, decreasing the number of interfaces  implies increasing the amount of superlattice that 
can be considered PTO bulk.
This results in a larger measured bias, assuming a constant density of defects in the bulk
regions.

\subsection{PTO/SRO superlattices}

In PTO/SRO superlattices, due to the lack of compositional inversion symmetry \cite{CalloriPRL2012} there are a total of 5 unique divacancy sites.
These are presented in the right model of Fig. \ref{fig:stosro_vacloc}.
In the figure we also indicate the direction of the
pristine superlattice intrinsic polarization for the corresponding atomic arrangement. 
This intrinsic polariation asymmetry, originates from the 
lack of compositional inversion symmetry \cite{CalloriPRL2012}.
The corresponding pristine double well potential is shown in the two bottom figures for
a  (PTO)\textsubscript{7}/(SRO)\textsubscript{1}  and a  (PTO)\textsubscript{10}/(SRO)\textsubscript{1} superlattice systems.
The plots shows how the intrinsic asymmetry, which stabilizes the naturally preferred polarization over the opposite one, decreases with increasing PTO concentration, eventually converging towards the fully symmetric double well \cite{CalloriPRL2012}.

For the two superlatice structures we modelled a $2\sqrt{2}\times2\sqrt{2}$ in plane supercell area with an in-plane lattice constant constrained to the STO equilibrium lattice constant of 3.87 \AA{}.
The divacancies were initialized in the same way as before.
For divacancies formed at the interface, there are three unique possible locations, which also uniquely determine the direction of the divacancy dipole moment. 
The three interfacial divacancies, shown in Fig. \ref{fig:stosro_vacloc}, are in the PTO unit cell closer to the SrO layer (\textit{PTO Interface I}), in the PTO unit cell closer the RuO\textsubscript{2} layer (\textit{PTO Interface II}) and in the PbO-RuO\textsubscript{2} layers, or equivalently, in the PbRuO\textsubscript{3} (PRO) unit cell (\textit{PRO Interface}).
We note that divacancy dipole moments in \textit{PTO Interface I} and \textit{PTO Interface II} are parallel with the direction of the naturally preferred polarization, caused by the compositional asymmetry at the interfaces of the superlattice.
However, divacancy dipole moments formed in the \textit{PRO Interface} are antiparallel to the naturally preferred polarization.
For divacancies formed in the PTO central region, we must consider both orientations, because, in contrast to divacancies in PTO/STO, they are not symmetric under permutation due to the compositional inversion symmetry breaking.

We found that for all divacancies, except the one formed in the \textit{PRO Interface}, we were only able to stabilize the naturally preferred polarization state, i.e. the double well potential becomes a single well
and the system under those conditions can be characterized as a polar insulator\cite{Callori2012}. 
This occurs because the system poled on the opposite polarization to that of the divacancy relaxes back into
the naturally preferred polarization state for the considered volume fractions.
However, something interesting occurs when the divacancy is located at the \textit{PRO Interface}.
In this case, the two nonequivalent polarizations can be stabilized.
In the 7/1 volume fraction system, the naturally preferred polarization (pointing down) is still the global minimum,
despite the dipole moment of the defect opposing it, as shown in Fig. \ref{fig:pto7sro1_dw}. 
However, the energy difference between the metastable state, i.e. the higher energy minimum of the 
asymmetric double well potential, and the stable state decreases by $\sim$ 1 eV  relative to the pristine double well. 
This occurs because the defect dipole moment in the metastable state is aligned with the system polarization, but anti-aligned with the naturally preferred one, which still dominates the overall polarization double well asymmetry.

With larger PTO volume fractions, in the 10/1 superlattice, we observe a change of behavior.
In this case a dipolar defect in the \textit{PRO Interface} is sufficient to reverse the preferred polarization direction so that it is now in the opposite direction to the polarization preference produced by the compositional symmetry breaking, as shown in Fig. \ref{fig:pto10sro1_dw}. 
We have considered all possible defect locations, and their corresponding results are presented in Supplemental section V.
Only when the defect is located at the \textit{PRO Interface} we observe a stable polarization sign change
with increasing PTO volume fraction.
These results would explain the experimental observations, which show how the bias in PTO/SRO  superlattices
presents a sign change with increasing PTO concentration.
The bias reversal can be understood as a compositionally  dominated bias at low volume fractions and  defect dominated bias at high volume fractions.

These findings would explain the experimental results if there was a strong preference for the development of defect dipoles in the \textit{PRO Interface} over other possible sites.
However, we note that the configuration where the defect is located in the \textit{PRO Interface} is 0.97 eV higher in energy relative to when the defect is located in the PTO layers.
We have addressed this inconsistency through a calculation of the stability of divacancies, presented in the supplemental Fig. S8.
The main discrepancy between our simulations and experiments is
that, due to computational limitations, the defect concentrations 
considered are 1-2 orders of magnitude larger than standard experimental values.
The results of our modeling of how electrostatic boundary conditions influence the stability of the divacancies in Supplemental Section S.V(A) show that the stability of the defect in the \textit{PRO Interface} increases with decreasing defect concentration.
%


\section{Experimental design of built-in bias free superlattices}

\begin{figure}
    \centering
    \includegraphics[width=8.5cm]{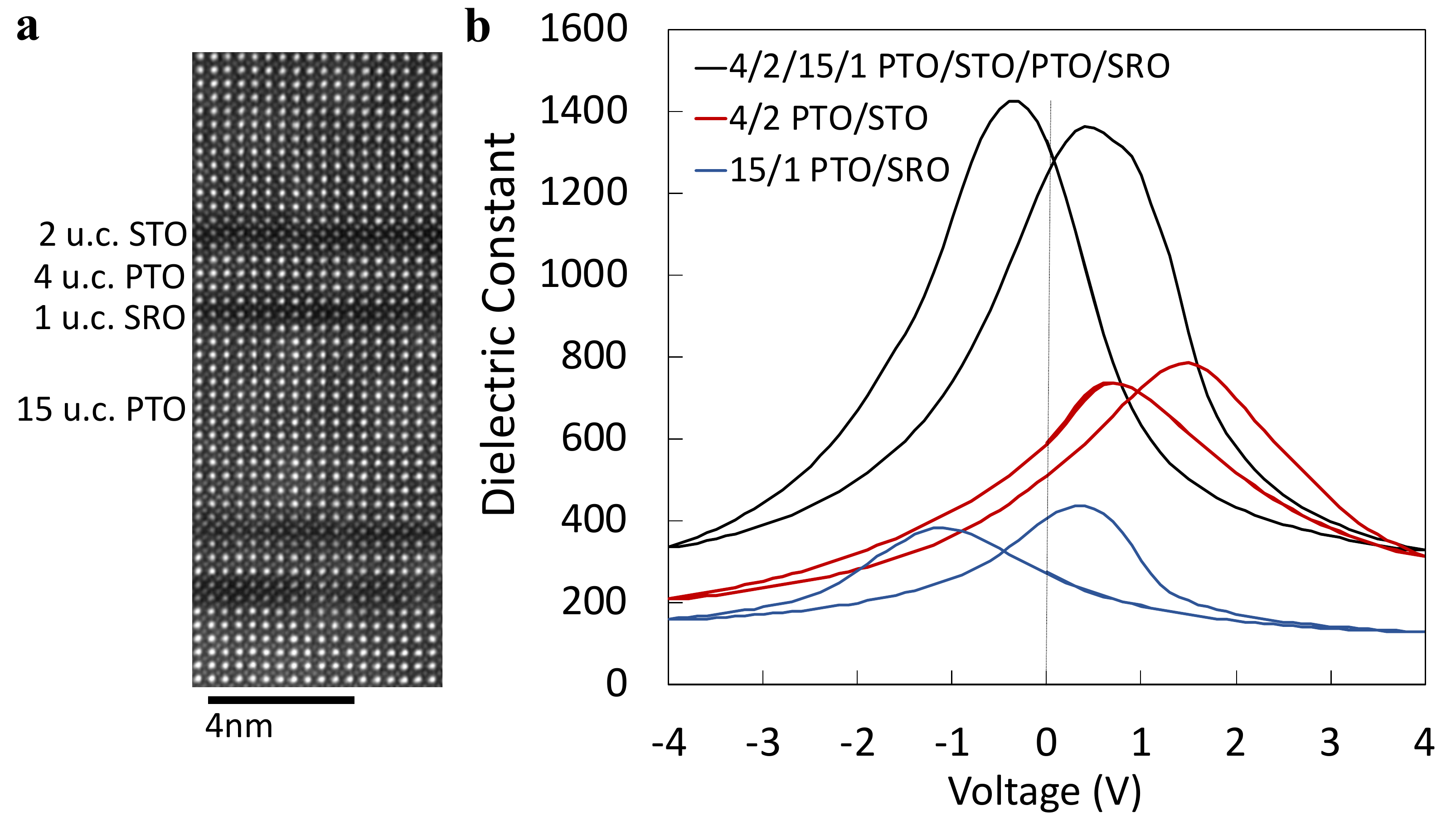}
    \caption{A hybrid superlattice that combines PbTiO\textsubscript{3}/SrTiO\textsubscript{3}
with PbTiO\textsubscript{3}/SrRuO\textsubscript{3}.(a)  A HRTEM cross-sectional image of a 15/1/4/2 PTO/STO/PTO/SRO
superlattice. (b) The  CV butterfly loop of the hybrid superlattice
sample and the two building block superlattices that were combined to make it}
\label{combo}
\end{figure}

The previous sections have used a combination of experimental and theoretical data to 
explain the microscopic origin of the observed built-in
bias in PTO/STO and PTO/SRO superlattices.
One practical consequence of understanding the bias origin is that if our
insights are correct, we could design bias free samples, not
by preventing the formation of defects, but by taking advantage of their presence.
This is facilitated by the differing signs of defect-induced bias in these two families of superlattices. 
Using as a guide the bias offset versus composition
of the PTO/STO superlattices and the PTO/SRO superlattices plotted
in Fig \ref{datasummary}, we produced a demonstration (n\textsubscript{1}/n\textsubscript{2}/n\textsubscript{3}/n\textsubscript{4})PTO/STO/PTO/SRO
combination superlattice, as illustrated in Fig. \ref{combo} (a). If the biases from one bilayer add simply those from another then with the correct n\textsubscript{3}/n\textsubscript{2} PTO/STO and
n\textsubscript{1}/n\textsubscript{4} PTO/SRO combinations, the total
sum of the built-in bias will average out and be near 0 V. Fig. \ref{combo}(a) is a cross-sectional HRTEM image of a (15/1/4/2)PTO/STO/PTO/SRO
sample showing good epitaxial growth and layering of just such a hybrid superlattice.  In Fig \ref{combo}(b) C-V measurements show good ferroelectric switching as well as a built-in bias near zero. The fact that the built-in bias of the PTO/STO and PTO/SRO bilayers can effectively cancel is strong evidence
for our argument that the alignment of Pb-O divacancy dipoles is determined as each layer is grown.

\section{Conclusions}
This study has shown that interfacial Pb-O divacancies in PbTiO$_3$-based
superlattices produce a large built-in bias, which depends on the composition of the 
supelattices.
We have correlated the position and orientation of the divacancies in the superlattice with
the observed bias using a combination
of experimental and first-principles simulation results.
We have also shown that the divacancy-induced built-in bias co-exists with  other sources of symmetry breaking such as compositional inversion symmetry breaking, which is present in PbTiO$_3$/SrRuO$_3$ superlattices.
While in superlattices without inversion symmetry breaking the bias always has the same sign, in
PbTiO$_3$/SrRuO$_3$ superlattices the sign of the bias depends on the concentration of PbTiO$_3$
relative to SrRuO$_3$.
The origin of this dependency is linked to the coupling between the dipole moment of the divacancies and the ferroelectric
polarization in the material, which in turn is uniquely fixed
by the superlattice growth.
Finally, we demonstrate that it is possible to 
engineer bias-free systems without having to eliminate the defects in the oxide superlattices, giving significantly more freedom in the space of experimental parameters that can be used for growth. 
\section{Acknowledgements}
We acknowledge funding from the National Science Foundation DMR-1334867 and DMR-1334428. SD and MVF-S. thank Stony Brook Research Computing and Cyberinfrastructure, and the Institute for Advanced Computational Science at Stony Brook University for access to the high-performance SeaWulf computing system, which was made possible by a \$1.4M National Science Foundation grant (\#1531492). We acknowledge very helpful discussions with Onur Erten and Cyrus Dreyer.

\bibliography{main}
\end{document}